\documentclass[aps,prd,twocolumn,preprintnumbers,showpacs,%
superscriptaddress,nofootinbib,amsmath,amssymb]{revtex4}

\begin{document}

\preprint{arXiv:0904.0668 [hep-th]}

\title{Bending AdS Waves with New Massive Gravity}

\author{Eloy Ay\'on--Beato}
\email{ayon-beato-at-fis.cinvestav.mx}
\affiliation{Departamento~de~F\'{\i}sica,~CINVESTAV--IPN,%
~Apdo.~Postal~14--740,~07000,~M\'exico~D.F.,~M\'exico.}

\author{Gaston Giribet}
\email{gaston-at-df.uba.ar} \affiliation{Departamento de
F\'{\i}sica, Universidad de Buenos Aires FCEN - UBA and CONICET,
Argentina.} \affiliation{Abdus Salam International Centre for
Theoretical Physics ICTP, Trieste, Italy.}

\author{Mokhtar Hassa\"{\i}ne}
\email{hassaine-at-inst-mat.utalca.cl} \affiliation{Instituto de
Matem\'atica y F\'{\i}sica, Universidad de Talca, Casilla 747,
Talca, Chile.}

\begin{abstract}
We study AdS-waves in the three-dimensional {\emph new} theory
of massive gravity recently proposed by Bergshoeff, Hohm, and
Townsend. The general configuration of this type is derived and
shown to exhibit different branches, with different asymptotic
behaviors. In particular, for the special fine tuning
$m^2=\pm1/(2l^2)$, solutions with logarithmic fall-off arise,
while in the range $m^2>-1/(2l^2)$, spacetimes with
Schr\"odinger isometry group are admitted as solutions.
Spacetimes that are asymptotically AdS$_3$, both for the
Brown-Henneaux and for the weakened boundary conditions, are
also identified. The metric function that characterizes the
profile of the AdS-wave behaves as a massive excitation on the
spacetime, with an effective mass given by
$m_{\mathrm{eff}}^2=m^2-1/(2l^2)$. For the critical value
$m^2=-1/(2l^2)$, the value of the effective mass precisely
saturates the Breitenlohner-Freedman bound for the AdS$_3$
space where the wave is propagating on. The analogies with the
AdS-wave solutions of topologically massive gravity are also
discussed. Besides, we consider the coupling of both massive
deformations to Einstein gravity and find the exact
configurations for the complete theory, discussing all the
different branches exhaustively. One of the effects of
introducing the Chern-Simons gravitational term is that of
breaking the degeneracy in the effective mass of the generic
modes of pure New Massive Gravity, producing a fine structure
due to parity violation. Another effect is that the zoo of
exact logarithmic specimens becomes considerably enlarged.
\end{abstract}

\maketitle

\section{Introduction}

A new theory of massive gravity in three dimensions has been
recently proposed by Bergshoeff, Hohm, and Townsend
\cite{Bergshoeff:2009hq}. At the linearized level, this theory has
been shown to be equivalent to the three-dimensional Fierz-Pauli
action for a massive spin-2 field, which turns out to be unitary. In
contrast with the Topologically Massive Gravity
\cite{Deser:1981wh,Deser:1982vy}, this new theory of gravity is
parity invariant. Following the authors of
\cite{Clement:2009gq,Liu:2009bk,Liu:2009kc}, we will refer to this
theory as the New Massive Gravity. As it is the case for
Topologically Massive Gravity, the New Massive Gravity entails
higher order modifications to the three-dimensional General
Relativity with the consequence that the graviton excitations of
both theories become massive through similar mechanisms.

This New theory of Massive Gravity has attracted much attention
recently. In Refs.~\cite{Nakasone:2009bn,Nakasone:2009vt} the
unitarity of the model was discussed; in Ref.~\cite{Clement:2009gq}
exact solutions representing asymptotically warped-AdS$_3$ black
holes were found; other interesting solutions are analyzed in
\cite{Julio}. In Refs.~\cite{Liu:2009bk,Liu:2009kc}, the boundary
conditions for the theory in asymptotically AdS$_3$ spaces were
studied. In particular, it was argued in Ref.~\cite{Liu:2009kc}
that, for a particular relation between the cosmological constant
and the mass parameter of the theory, the concept of asymptotically
AdS$_3$ boundary conditions is compatible with a weakened
logarithmic fall-off at large distance. This amounts to relax the
standard Brown-Henneaux asymptotic conditions for gravity in AdS$_3$
space \cite{Brown:1986nw}, and it is analogous to what happens in
Topologically Massive Gravity at the chiral point
\cite{Grumiller:2008es,Henneaux:2009pw}. In the case of
Topologically Massive Gravity, the feasibility of considering
weakened AdS$_3$ boundary conditions allows to reinterpret some of
the AdS-waves with logarithmic profile found in
Refs.~\cite{AyonBeato:2004fq,AyonBeato:2005qq} as being
asymptotically AdS$_3$ spacetimes. These configurations, together
with those that do not exhibit the logarithmic decay,\footnote{See
Ref.~\cite{Dereli:2000fm} for a preliminary derivation of the
no-logarithmic branch, where it was argued that in order to be
supersymmetric the solutions should not depend on the retarded time.
See also Ref.~\cite{Olmez:2005by}, where the interpretation as
AdS-waves was first given to the final forms of the metric
originally derived in Ref.~\cite{AyonBeato:2004fq}.} have been
reconsidered recently \cite{Carlip:2008eq,Gibbons:2008vi} within the
context of the revived interest in Topologically Massive Gravity
\cite{Li:2008dq,chirales,warped,Yerko,Stromingernew}.

Here, we will explore the AdS-wave configurations of the New Massive
Gravity. In particular, this will allow us to study the dynamics of
the theory beyond the linearized level and to discuss the different
asymptotic behaviors; a strategy proved to be useful in
Topologically Massive Gravity \cite{AyonBeato:2005qq}. For example,
we will find that solutions with logarithmic fall-off arise for
certain values of the parameters. The existence of such solutions to
New Massive Gravity is suggested by the linearized analysis
performed in Ref.~\cite{Liu:2009kc}. Nevertheless, in this paper we
go beyond the perturbative analysis and find exact solutions of this
kind. More generally, we will exhibit a whole family of AdS-wave
solutions of the New Massive Gravity; and these solutions are, in
some sense, the analogues of those arising in Topologically Massive
Gravity
\cite{AyonBeato:2004fq,AyonBeato:2005qq,Dereli:2000fm,Olmez:2005by,%
Carlip:2008eq,Gibbons:2008vi}. For a particular range of the
parameters, we will show that the isometry of the solutions
coincides with the Schr\"odinger symmetry. We will also analyze
similar considerations for a more general theory given by the
coupling between both massive gravity theories.

The paper is organized as follows. In Sec.~\ref{sec:introAdSw}, we
provide a brief introduction to AdS-wave configurations. Section
\ref{sec:AdSwNMG} is devoted to present the New Massive Gravity of
Ref.~\cite{Bergshoeff:2009hq}. The general AdS-wave solutions are
derived and the different branches depending on their asymptotic
behaviors are analyzed. In Sec.~\ref{sec:analgTMG}, we discuss some
analogies between these configurations and their cousins arising in
Topologically Massive Gravity; we also point out some differences.
Besides, we consider the coupling between both massive models and
derive the corresponding AdS-wave configurations. The nontrivial
effects due to the inclusion of the topological mass term are
analyzed in details in Sec.~\ref{sec:AdSwNMGC}.

\section{Brief Introduction to AdS waves\label{sec:introAdSw}}

AdS waves are a special kind of exact gravitational waves
propagating along AdS space. The first examples of exact
gravitational waves in the presence of a cosmological constant
were studied by Garc\'{\i}a and Pleba\'nski \cite{Garcia:1981};
see also Refs.~\cite{Salazar:1983,Garcia:1983,Ozsvath:1985qn}.
Such solutions were based on generalizations of some
algebraically special solutions previously found for the case
of vanishing cosmological constant in
Refs.~\cite{Kundt:1961,Robinson:1960}. The algebraically
special spacetimes are defined by the fact that their Weyl
tensor has a multiple principal null direction. In addition, if
this null direction is a Killing vector for the exact wave
solutions then, in the case of a negative cosmological
constant, one recovers the so-called Siklos spacetimes defining
the AdS waves \cite{Siklos:1985}. Siklos spacetimes allow an
alternative characterization as a generalized Kerr-Schild
transformation of AdS, which reinforces their interpretation as
exact gravitational waves propagating on AdS space
\cite{Podolsky:1997ik}. This means that their metrics can be
written in terms of the AdS metric as follows
\begin{equation}\label{eq:K-S}
g_{\mu\nu}=g^{\mathrm{AdS}}_{\mu\nu}-Fk_{\mu}k_\nu,
\end{equation}
where $k^\mu$ is a null geodesic field, and $F$ is an arbitrary
function that is only constrained to be independent of the integral
parameter along $k^\mu$; see Ref.~\cite{AyonBeato:2005qq} for a more
detailed discussion.

Now, let us consider the three-dimensional case we are interested
in. The AdS$_3$ metric in Poincar\'e coordinates reads
\begin{equation}\label{eq:AdS}
ds^2_{\mathrm{AdS}}=\frac{l^2}{y^2}\left(-2dudv+dy^2\right),
\end{equation}
where $l$ is the radius of AdS characterizing its constant scalar
curvature $R=6\Lambda=-6/l^2$. Choosing as null geodesic field
$k^\mu\partial_\mu=(y/l)\partial_v$, the Kerr-Schild transformation
(\ref{eq:K-S}) allows to write the metric of the AdS$_3$-waves as
follows
\begin{equation}\label{eq:AdSwave}
ds^2=\frac{l^2}{y^2}\left[-F(u,y)du^2-2dudv+dy^2\right].
\end{equation}

This metric is conformally related to that of a \emph{pp}-wave.
Nevertheless, it is worth pointing out that AdS-waves and
\emph{pp}-waves have different geometrical and physical properties.
Let us be reminded of the fact that the term \emph{pp}-wave stands
for \emph{p}lane fronted gravitational waves with \emph{p}arallel
rays. The fronts of the wave are defined by surfaces
$u,v=\mathrm{const.}$ in any number of dimension, and for the
AdS-waves in higher dimensions they are not \emph{p}lanes but
hyperboloids, having constant curvature proportional to $-1/l^2$.
Additionally, the null rays are defined by the field $\partial_v$,
which is a Killing vector but not a closed 1-form, and thus the rays
fail to be covariantly constant, namely \emph{p}arallel.

We will explore the existence of AdS waves configurations
rigged by the New Massive Gravity of
Ref.~\cite{Bergshoeff:2009hq}.

\section{The AdS waves of New Massive Gravity\label{sec:AdSwNMG}}

The action of the New Massive Gravity is\footnote{We follow the
conventions of \cite{Liu:2009kc}.}
\begin{equation}\label{eq:S}
S=\frac{1}{16\pi G}\int{d}^3x\sqrt{-g}
\left(R-2\lambda-\frac1{m^2}K\right),
\end{equation}
where the quadratic contribution $K=R_{\mu\nu}R^{\mu\nu}-\frac38R^2$
introduces the modification to standard gravity with cosmological
constant $\lambda$, being $m$ the mass of the resulting massive
degrees of freedom. The variation of (\ref{eq:S}) gives rise to the
modified gravity equations of motion
\begin{equation}\label{eq:NMG}
G_{\mu\nu}+\lambda{g}_{\mu\nu}-\frac1{2m^2}K_{\mu\nu}=0,
\end{equation}
where $G_{\mu\nu}=R_{\mu\nu}-\frac12g_{\mu\nu}R$ is the Einstein
tensor and
\begin{eqnarray}\nonumber
K_{\mu\nu}&=&2\square{R}_{\mu\nu}-\frac12\nabla_\mu\nabla_\nu{R}
-\frac12\square{R}g_{\mu\nu} +4R_{\mu\alpha\nu\beta}R^{\alpha\beta}\\
&&{}-\frac32RR_{\mu\nu}-Kg_{\mu\nu},
\end{eqnarray}
is a symmetric, conserved tensor that satisfies
$g^{\mu\nu}K_{\mu\nu}=K$. This condition on $K_{\mu\nu}$ implies
that the trace of the equations of motion is a second order
constraint, despite the fact these are equations of fourth order.

\subsection{AdS-waves solutions}

For the New Massive Gravity to admit an AdS$_3$ vacuum
(\ref{eq:AdS}), a special constraint between the AdS$_3$ radius $l$,
the cosmological constant $\lambda$, and the mass parameter $m$ is
needed. This fixes the value of the cosmological constant to be
\cite{Bergshoeff:2009hq,Liu:2009kc}
\begin{equation}\label{eq:lambda}
\lambda=-\frac1{l^2}\left(1+\frac1{4l^2m^2}\right),
\end{equation}
which means the scale of the cosmological constant and the AdS
radius only coincide in the General Relativity limit
$m^2\to\pm\infty$.

The AdS-wave solutions (\ref{eq:AdSwave}) are meant to describe
exact gravitational waves propagating along AdS$_3$ spacetime of
radius $l$, and thus we have to consider the same election
(\ref{eq:lambda}) for the cosmological constant. With this choice
for $\lambda$, the equations of motion (\ref{eq:NMG}) become a
single differential equation for the wave profile $F$; namely
\begin{eqnarray}
&&\bigg[y^4\partial^4_yF+2y^3\partial^3_yF\nonumber\\
&&\quad{}-\frac{(1+2l^2m^2)}2
\left(y^2\partial^2_yF-y\partial_yF\right)\bigg]
\frac{\delta_{\mu}^u\delta_{\nu}^u}{2l^2m^2y^2}=0,\qquad
\label{eq:NMGAdSwave}
\end{eqnarray}

This is a fourth order Euler-Fuchs differential equation, which is
easily solved by applying the standard substitution $F=y^\alpha$.
The corresponding fourth-degree characteristic polynomial is
\begin{equation}\label{eq:charpol}
\alpha(\alpha-2)\left((\alpha-1)^2-\frac{1+2l^2m^2}2\right)=0.
\end{equation}
Therefore, the {\it generic} solution for the wave profile is
\begin{equation}\label{eq:Fgen}
F(u,y)=F_+(u)\left(\frac{y}l\right)^{1+\sqrt{\frac{1+2l^2m^2}2}}+
F_-(u)\left(\frac{y}l\right)^{1-\sqrt{\frac{1+2l^2m^2}2}},
\end{equation}
where $F_+$ and $F_-$ are arbitrary integration functions that
depend only on the retarded time $u$. Here and in what follows, we
also use the fact that the homogeneous and quadratic dependence of
the wave-front coordinate $y$ can be eliminated by coordinate
transformations, see the detailed discussion in
Ref.~\cite{AyonBeato:2005qq}.

In addition to (\ref{eq:Fgen}), we have to consider the possibility
of having multiplicities in the roots of the characteristic
polynomial (\ref{eq:charpol}). In this case, the power-law
particular solutions fail to span the whole space of linearly
independent solutions, and thus new additional logarithmic modes
appear. Such multiplicities arise for the mass values
$m^2=\pm1/(2l^2)$. For $m^2=-1/(2l^2)$, there exists double
multiplicity; the two roots exhibited in the generic solution
(\ref{eq:Fgen}) become one. Then, after discarding trivial
behaviors, the wave profile at this point turns out to be given by
\begin{equation}\label{eq:Fm2=-1/2l2}
F(u,y)=\frac{y}l\left[F_1(u)\ln\left(\frac{y}l\right)+
F_2(u)\right].
\end{equation}

On the other hand, for $m^2=+1/(2l^2)$ we find double multiplicity
both for $\alpha=0$ and for $\alpha=2$, because in this case the
roots of the generic solution (\ref{eq:Fgen}) reduce to these
values. Then, in this case we are left with the following solution
\begin{equation}\label{eq:Fm2=+1/2l2}
F(u,y)=\ln\left(\frac{y}l\right)\left[F_1(u)\left(\frac{y}l\right)^2
+F_2(u)\right].
\end{equation}

Finally, for $m^2<-1/(2l^2)$ the relevant roots of
(\ref{eq:charpol}) take complex values, and the solution becomes
\begin{eqnarray}
F(u,y)&=&\frac{y}l\left\{ F_1(u)\sin\left[l\sqrt{-\frac1{2l^2}-m^2}
\ln\left(\frac{y}l\right)\right]\right.\nonumber\\
&&\left.{}+F_2(u)\cos\left[l\sqrt{-\frac1{2l^2}-m^2}
\ln\left(\frac{y}l\right)\right]\right\}.\qquad\label{eq:Fm2<-1/2l2}
\end{eqnarray}

The configurations given by (\ref{eq:Fgen})-(\ref{eq:Fm2<-1/2l2})
represent the AdS$_3$-wave solutions to the New theory of Massive
Gravity \cite{Bergshoeff:2009hq}. In Section IV we will discuss the
analogy between these solutions and those arising in the context of
the Topologically Massive Gravity. But, first, let us comment on the
asymptotic behavior of the solutions we just described.

\subsection{The asymptotically AdS$_3$ sector}

As mentioned, AdS-waves are Siklos spacetimes \cite{Siklos:1985}
that can be thought of as gravitational wave profiles propagating on
AdS spacetime \cite{Podolsky:1997ik}. Here, we will show that, in
addition, some of these wave solutions of New Massive Gravity are
also asymptotically AdS$_3$.

New Massive Gravity in AdS$_3$ has been recently studied in
\cite{Bergshoeff:2009hq,Liu:2009bk,Liu:2009kc}. According to
AdS$_3$/CFT$_2$ correspondence, the theory formulated in AdS$_3$
would be dual to a two-dimensional conformal field theory with
central charge given by
\begin{equation}\label{eq:centralcharge}
c= \frac{3l}{2G}\left(1-\frac1{2m^2l^2}\right).
\end{equation}
This value for the central charge is easily obtained by standard
means \cite{Kraus}. From this we observe that something special
happens at $m^2=1/(2l^2)$, where $c$ vanishes. Likely, the unitarity
of the theory (when sufficiently relaxed boundary conditions are
considered) would demand the bound $m^2>1/(2l^2)$. Let us discuss
the different asymptotic behaviors in relation to this bound.

First, let us take a look at solutions (\ref{eq:Fgen}). The wave solution turns out to be an asymptotically
AdS$_3$ spacetime if $F_-= 0$ and $m^2>1/(2l^2)$. That is, the solution is asymptotically AdS$_3$ according to Brown-Henneaux
boundary conditions \cite{Brown:1986nw}, which, in these
coordinates, are defined by the next-to-leading behavior
\begin{equation}\label{eq:expansion}
g_{\mu\nu} = g_{\mu\nu}^{\mathrm{AdS}} + h_{\mu\nu},
\end{equation}
where $g_{\mu\nu}^{\mathrm{AdS}}$ is given by (\ref{eq:AdS}), while
the components of the perturbation $h_{\mu\nu}$ are of order
$h_{uu}\sim h_{uv} \sim h_{vv}\sim h_{yy}\sim\mathcal{O}(1)$, and
$h_{uy}\sim h_{vy}\sim\mathcal{O}(y)$.

On the other hand, at the critical value $m^2=1/(2l^2)$, solution
(\ref{eq:Fm2=+1/2l2}) turns out to be compatible with the weakened
(logarithmic) AdS$_3$ asymptotic behavior discussed in
Refs.~\cite{Grumiller:2008es,Henneaux:2009pw,Yerko,Stromingernew},
which amount to relax boundary conditions as
$h_{uu}\sim\mathcal{O}(\ln{y})$ and
$h_{uy}\sim\mathcal{O}(y\ln{y})$. These weakened AdS$_3$ boundary
conditions were originally discussed within the context of
Topologically Massive Gravity, where the analog of solution
(\ref{eq:Fm2=+1/2l2}) given in
Refs.~\cite{AyonBeato:2004fq,AyonBeato:2005qq} is
Eq.~(\ref{eq:Fmu=-1/lTMG}) below, and it was recently argued that
they might play an important role in New Massive Gravity too
\cite{Liu:2009kc}. It has been known for a while that, for certain
particular points of the space of parameters of a given theory, the
concept of asymptotically AdS$_3$ space may be consistently extended
to incorporate a larger class of geometries \cite{HMTZ}. This issue
has played an important role in recent discussions on Topologically
Massive Gravity \cite{Stromingernew}.

\section{Analogies with Topologically Massive Gravity
\label{sec:analgTMG}}

It is interesting to notice that all the branches of solutions
discussed above, except the complex one (\ref{eq:Fm2<-1/2l2}), have
their counterparts in Topologically Massive Gravity
\cite{AyonBeato:2004fq,AyonBeato:2005qq,Dereli:2000fm,Olmez:2005by,%
Carlip:2008eq,Gibbons:2008vi}. In particular, the critical cases
$m^2=\pm1/(2l^2)$ deserve a particular attention because these are
reminiscent of the chiral values $\mu=\pm1/l$ of Topologically
Massive Gravity, with $\mu$ being the topological mass. These points
of the space of parameters were shown to be special in what regards
to the massive behavior of AdS$_3$-wave solutions
\cite{AyonBeato:2005qq}. More recently, the points $\mu=\pm1/l$
appeared to be relevant also for the discussion about the chiral
gravity conjecture \cite{Li:2008dq}; see also
\cite{Grumiller:2008es,Carlip:2008eq,chirales,Stromingernew} and
references therein.

The purpose of this section is to discuss this and other
analogies between the AdS-wave solutions of New Massive Gravity
and those of Topologically Massive Gravity.

\subsection{The Schr\"odinger invariant sector}

Recently, a generalization of the AdS/CFT correspondence has been
proposed in the context of non-relativistic conformal field
theories. The basic idea is that geometries whose isometry group
agrees with the non-relativistic conformal group, namely the
Schr\"odinger group, could represent gravity duals for systems of
condensed matter physics \cite{Son:2008ye,Balasubramanian:2008dm}.
The Schr\"odinger group is defined as the maximal group of
symmetries which leave invariant the Schr\"odinger equation for a
free particle \cite{JackiwPT,Schr}, and can be thought of as the
semi direct product of $\mathop{\rm SL}(2,\mathbb{R})$ with the
connected static Galilei group. The set of Schr\"odinger
transformations are given by the standard Galilei transformations
augmented by the time dilatation and a \emph{special} conformal
transformation.

Because of the holographic applications to non-relativistic CFTs,
the search of theories that admit as solutions
Schr\"odinger-invariant backgrounds has attracted much attention
recently. Moreover, it is worth mentioning that the AdS-waves of
Topologically Massive Gravity \cite{AyonBeato:2005qq} contains
Schr\"odinger invariant solutions at the special point $\mu=3/l$
\cite{Duval:2008jg}. These solutions correspond to the null
warped-AdS$_3$ spacetimes of \cite{warped}. Here, let us show that
our generic solution (\ref{eq:Fgen}) of New Massive Gravity may
exhibit the Schr\"odinger isometry too. In fact, if one takes one of
the arbitrary functions $F_{\pm}$ to be a constant and the other one
to be zero, then the solution (\ref{eq:Fgen}) simply reads
\begin{equation}\label{eq:asterisc}
F(y) = F_0 \left(\frac{y}{l}\right)^{-2\nu}
\end{equation}
where $F_0$ is an arbitrary constant and $\nu=-\alpha/2$ is called
the ``dynamical exponent'' in this context. Interesting enough, the
isometry group of metric (\ref{eq:AdSwave}) for solution
(\ref{eq:asterisc}) gets enhanced, exhibiting the so-called
\emph{partial} Schr\"odinger group, which is the group of all
Schr\"odinger transformations except the {\it special} conformal
transformation. The \emph{partial} Schr\"odinger symmetry is
realized by the Killing vectors
\begin{eqnarray*}
H &=&\partial_v, \qquad N=\partial_u, \\
D &=&(1+\nu)v\partial_v+(1-\nu)u\partial_u+y\partial_y.
\end{eqnarray*}
In addition, the particular election $F_+=0$ and
$F_-=\mathrm{const.}$ allows the special case $\nu=1$, which
corresponds to
\begin{eqnarray}
m^2=\frac{17}{2l^2}, \label{ftSchro}
\end{eqnarray}
and the solution exhibits the \emph{full} Schr\"odinger symmetry,
i.e.\ the isometry group is augmented by the Killing vector
\[
C=v^2\partial_v+\frac12y^2\partial_u+yv\partial_y.
\]

This critical point (\ref{ftSchro}) is analogous to the point
$\mu=3/l$ of the Topologically Massive Gravity (see
Eq.~(\ref{eq:FgenTMG}) below). This reinforces the resemblance
between both theories.

\subsection{AdS waves as massive scalar modes: log waves
saturate the BF bound}

Another interesting property of solutions (\ref{eq:Fgen}) is that
the profile function $F$ behaves exactly as a massive scalar mode,
as it satisfies the Klein-Gordon equation
\begin{equation}\label{eq:K_Gg}
\Box{F}={m^2_{\mathrm{eff}}} F,
\end{equation}
with effective mass given by
\begin{equation}\label{eq:m_eff}
m_{\mathrm{eff}}^2=m^2-\frac1{2l^2}.
\end{equation}
This shifting of the \emph{bare} mass $m$ by a term proportional to
the curvature of the AdS$_3$ space is also observed in the case of
AdS$_3$-waves of Topologically Massive Gravity, where the effective
mass is found to be
\[
\mu^2_{\mathrm{eff}} = \mu^2 -\frac{1}{l^2},
\]
(see Ref.~\cite{AyonBeato:2005qq} for details). However, the
New Massive Gravity profile (\ref{eq:Fgen}) describes in fact
the superposition of two scalar modes, in contrast with
Topologically Massive Gravity, for which a single mode arises
(see Eq.~(\ref{eq:FgenTMG}) later). Actually, at this level it
may seem artificial to make a distinction between the ``two
modes'' appearing in Eq.~(\ref{eq:Fgen}) since, after all, they
have the same effective mass (\ref{eq:m_eff}). However, we will
see in the next section that the inclusion of a topologically
massive term (i.e.\ the Chern-Simons gravitational term) breaks
this mass degeneracy and thus the distinction between the two
modes ultimately makes sense.

In the case of the New Massive Gravity, and for the special case
$m^2=-1/(2l^2)$, the profile (\ref{eq:Fm2=-1/2l2}) also describes
the superposition of two exact massive scalar modes, each one
satisfying
\begin{equation}\label{eq:K_Gm2=BFb}
\Box{F}=-\frac1{l^2}F.
\end{equation}
This case does not differ from the generic one (\ref{eq:K_Gg}) since
the effective mass becomes $m_{\mathrm{eff}}^2=-1/l^2$ and thus
corresponds to the mass given by (\ref{eq:m_eff}). Interesting
enough, we find that this value for the effective mass exactly
saturates the Breitenlohner-Freedman bound for the mass of a scalar
field in the AdS$_3$ space where the wave is propagating on
\cite{Breitenlohner:1982bm,Mezincescu:1984ev}.

The result (\ref{eq:K_Gm2=BFb}) for the case $m^2=-1/(2l^2)$ is in
contrast with what happens in Topologically Massive Gravity, where
none of the logarithmic solutions that appear at $\mu=\pm1/l$ (see
Eqs.~(\ref{eq:Fmu=+1/lTMG}) and (\ref{eq:Fmu=-1/lTMG}) below)
satisfy the Klein-Gordon equation. The analogy with Topologically
Massive Gravity is thus manifested at the other critical point,
$m^2=+1/(2l^2)$. At this point, the wave profile
(\ref{eq:Fm2=+1/2l2}) does not satisfy a Klein-Gordon equation, and,
as mentioned before, this is precisely the point of the space of
parameters where the asymptotically AdS$_3$ spaces admit logarithmic
branches \cite{Liu:2009kc} similar to those appearing for
Topologically Massive Gravity at the chiral point $\mu=-1/l$,
\cite{Grumiller:2008es,Henneaux:2009pw}\footnote{Our definition of
the topological mass is minus the one of those references, where the
chiral point occurs for $\mu=+1/l$.}. Also, it is interesting to
notice that the New Massive Gravity profile (\ref{eq:Fm2=+1/2l2}) is
exactly a superposition of the critical profiles
(\ref{eq:Fmu=+1/lTMG}) and (\ref{eq:Fmu=-1/lTMG}) of Topologically
Massive Gravity
\begin{equation}\label{eq:FNMGsuperpTMG}
F = \left.F_{\mathrm{TMG}}\right|_{\mu=-1/l} +
\left.F_{\mathrm{TMG}}\right|_{\mu=+1/l}.
\end{equation}
We remark that, despite the fact the profile function $F$ in
Topologically Massive Gravity at the points $\mu=\pm1/l$ do not obey
the Klein-Gordon equation, it can be used to generate the following
scalar modes,
\begin{equation}\label{eq:TMGc->sm}
\frac{l}y\left.F_{\mathrm{TMG}}\right|_{\mu=-1/l}, \qquad
\mathrm{and} \qquad
\frac{y}l\left.F_{\mathrm{TMG}}\right|_{\mu=+1/l},
\end{equation}
which do satisfy the Klein-Gordon equation (\ref{eq:K_Gm2=BFb})
saturating the Breitenlohner-Freedman bound. That is, one can
interpret the profile (\ref{eq:Fm2=+1/2l2}) of New Massive Gravity
at the critical point $m^2=+1/(2l^2)$ as a \emph{local}
superposition
\begin{equation}\label{eq:loc_superp}
F = \frac{y}l\bigg[\frac{l}yF|_{F_2=0}\bigg] +
\frac{l}y\bigg[\frac{y}lF|_{F_1=0}\bigg],
\end{equation}
of exact massive scalar modes (those between brackets) which
saturate the Breitenlohner-Freedman bound.

So far, we have discussed the analogies between New Massive Gravity
and Topologically Massive Gravity. Now, let us move to analyze what
happens when these two theories are brought together.

\section{Turning on a Topological Contribution\label{sec:AdSwNMGC}}

In this section, we analyze the effect of turning on the
Chern-Simons topological term in the gravitational action
(\ref{eq:S}), and see how it affects the existence and properties of
AdS-wave configurations we discussed so far. The inclusion of the
topologically massive term in the action amounts to add the Cotton
tensor\footnote{Here, $\eta_{\mu\alpha\beta}$ corresponds to the
volume 3-form, with $\eta_{uvy}=\sqrt{-g}$
($\eta^{uvy}=-1/\sqrt{-g}$).} \cite{Deser:1981wh,Deser:1982vy}
\begin{equation}\label{eq:Cotton}
C^{\mu\nu} = \eta^{\mu\alpha\beta}\nabla_\alpha
\left(R_\beta^{~\nu}-\frac14 R\,\delta_\beta^{~\nu}\right),
\end{equation}
to the equations of motion (\ref{eq:NMG}). The resulting field
equations read \cite{Bergshoeff:2009hq}
\begin{equation}\label{eq:NMGC}
G_{\mu\nu}+\lambda{g}_{\mu\nu}-\frac1{2m^2}K_{\mu\nu}
+\frac1{\mu}C_{\mu\nu}=0,
\end{equation}
where the coupling constant $\mu$ stands for the topological mass.
It is known that the Cotton tensor vanishes for constant curvature
configurations. Therefore, in order for the AdS$_3$ metric
(\ref{eq:AdS}) to be a solution of the generalized equations
(\ref{eq:NMGC}), the constraint between the cosmological constant
$\lambda$, the AdS radius $l$, and the mass $m$, must be exactly the
same as in Eq.~(\ref{eq:lambda}).

For an AdS-wave (\ref{eq:AdSwave}), the only nonvanishing
component of the Cotton tensor is $C_{uu}$, and it is
proportional to the third derivative of the wave profile with
respect to the wave-front coordinate $y$. The resulting single
equation is again of the Euler-Fuchs type; namely
\begin{eqnarray}
&&\bigg[y^4\partial^4_yF
+\left(2-\frac{lm^2}{\mu}\right)y^3\partial^3_yF\nonumber\\
&&\quad{}-\frac{(1+2l^2m^2)}2\left(y^2\partial^2_yF
-y\partial_yF\right)\bigg]
\frac{\delta_{\mu}^u\delta_{\nu}^u}{2l^2m^2y^2}=0.\qquad
\label{eq:NMGCAdSwave}
\end{eqnarray}
The characteristic polynomial is now given by
\begin{equation}\label{eq:charpolC}
\alpha(\alpha-2)\left[\left(\alpha-1-\frac{lm^2}{2\mu}\right)^2
-\frac{1+2l^2m^2}2-\frac{l^2m^4}{4\mu^2}\right]=0.
\end{equation}
Below, we will analyze all the possible solutions.

\subsection{Topological mass splitting}

According to (\ref{eq:charpolC}), the generic solution is given by
\begin{eqnarray}
F(u,y)&=&F_+(u)\left(\frac{y}l\right)^{1+\frac{lm^2}{2\mu}+
\sqrt{\frac{1+2l^2m^2}2+\frac{l^2m^4}{4\mu^2}}}\nonumber\\
&&{} +F_-(u)\left(\frac{y}l\right)^{1+\frac{lm^2}{2\mu}-
\sqrt{\frac{1+2l^2m^2}2+\frac{l^2m^4}{4\mu^2}}}.\quad
\label{eq:FgenC}
\end{eqnarray}
This generalizes the \emph{generic} profile of New Massive gravity
(\ref{eq:Fgen}). It represents the superposition of two exact scalar
modes, both satisfying a Klein-Gordon equation (\ref{eq:K_Gg}).
These modes are given by $F_-=0$ (resp.\ $F_+=0$) with effective
masses given respectively by
\begin{equation}\label{eq:m_effC}
m_{\mathrm{eff}\pm}^2=\left(\frac{m^2}{2\mu}
\pm\sqrt{\frac{m^4}{4\mu^2}+m^2+\frac1{2l^2}}\right)^2 -\frac1{l^2}.
\end{equation}
From this we observe that the physical effect of including the
topological term is that of breaking the degeneracy in the mass
spectrum (\ref{eq:m_eff}) for the two scalar modes of
\emph{pure} New Massive Gravity (\ref{eq:Fgen}). In other
words, the topological term produces the following mass
splitting between the two generic gravitational states
\begin{equation}\label{eq:massgap}
{\Delta}m_{\mathrm{eff}}^2
=\frac{2m^2}{\mu}\sqrt{\frac{m^4}{4\mu^2}+m^2+\frac1{2l^2}}.
\end{equation}
It is interesting that one has access to this fine structure effect
beyond the perturbative level.

The spacetime configurations (\ref{eq:FgenC}) also contain sectors
enjoying the \emph{partial} Schr\"odinger isometry as in
Eq.~(\ref{eq:asterisc}), while the \emph{full} Schr\"odinger
symmetry is exhibited this time for
\begin{equation}\label{eq:massSch}
m^2=\frac{17\mu}{2l^2\left(\mu-3/l\right)}.
\end{equation}

Notice that the generic formulas of Sec.~\ref{sec:AdSwNMG} are
obtained from the previous ones in the limit $\mu\to\pm\infty$. It
is also remarkable that the critical value for which the AdS-waves
solutions of Topologically Massive Gravity are Schr\"odinger
invariant, i.e.\ $\mu=3/l$, is also recovered from the mass
(\ref{eq:massSch}) in the limit $m^2\to\pm\infty$. Actually, in this
limit, one of the two modes in Eq.~(\ref{eq:FgenC})
diverges/vanishes and has no analogue in Topologically Massive
Gravity, while the other mode corresponds precisely to the single
generic mode that appears in Topologically Massive Gravity
\cite{AyonBeato:2004fq,AyonBeato:2005qq}; namely
\begin{equation}\label{eq:FgenTMG}
F_{\mathrm{TMG}}(u,y)=F_1(u)\left(\frac{y}l\right)^{1-l\mu}.
\end{equation}

This two-to-one correspondence between the modes of both theories is
a common feature of this transition and is easily understood by
taking into account that, unlike in New Massive Gravity, parity is
broken in Topologically Massive Gravity, and this fact forces the
latter theory to select only one of the two modes arising in the
former.

\subsection{Logarithmic Branches}

Now, let us discuss the cases allowing multiplicities in the roots
of the characteristic polynomial (\ref{eq:charpolC}). These cases
are those where logarithmic branches arise.

The roots corresponding to the generic solution (\ref{eq:FgenC})
reduce to a single one for the two following families of mass values
\begin{equation}\label{eq:mass_dm1C}
m^2=2\mu^2\left(-1\pm\sqrt{1-\frac1{2l^2\mu^2}}\right).
\end{equation}
This double multiplicity leads to two new families of solutions;
namely
\begin{eqnarray}
F(u,y)&=&\left(\frac{y}l\right)^{1
+l\mu\left[-1 \pm \sqrt{1-1/(2l^2\mu^2)}\right]}\nonumber\\
&&{}\times\left[F_1(u)\ln\left(\frac{y}l\right)+
F_2(u)\right].\label{eq:Fmass_dm1C}
\end{eqnarray}
For the upper sign, and taking the limit $\mu\to\pm\infty$, the mass
goes like
\begin{equation}\label{eq:2-NMG}
m^2=-\frac1{2l^2}+\mathcal{O}\left(\frac1{\mu^2}\right),
\end{equation}
and then one recovers the first critical solution
(\ref{eq:Fm2=-1/2l2}), studied in Sec.~\ref{sec:AdSwNMG}. The same
limit is divergent for the lower-sign since
\begin{equation}\label{eq:2AdSG}
m^2=-4\mu^2+\frac1{2l^2}+\mathcal{O}\left(\frac1{\mu^2}\right),
\end{equation}
Actually, this describes the transition to the standard AdS$_3$
Einstein gravity instead of to \emph{pure} New Massive Gravity. It
is known that AdS waves are locally trivial in this context
\cite{AyonBeato:2005qq}, and consequently, one mode vanishes and the
other diverges in the above limit.

For the lower-sign family, and for the topological mass taking the
value $\mu=3/(4l)$ (resp.\ $\mu=-3/(4l)$), the root $\alpha=0$
(resp.\ $\alpha=2$) becomes triple and then the solutions in these
cases respectively read
\begin{equation}\label{eq:Fmu=+3/4l_tmC}
F(u,y)=\ln\left(\frac{y}l\right)
\left[F_1(u)\ln\left(\frac{y}l\right)+ F_2(u)\right],
\end{equation}
and
\begin{equation}\label{eq:Fmu=-3/4l_tmC}
F(u,y)=\left(\frac{y}l\right)^2\ln\left(\frac{y}l\right)
\left[F_1(u)\ln\left(\frac{y}l\right)+ F_2(u)\right].
\end{equation}

Another critical value for the mass is given by
\begin{equation}\label{eq:mass_dm2C}
m^2=\frac{\mu}{2l^2(\mu-1/l)},
\end{equation}
where the value $\alpha=0$ turns out to be a double root. The
solution, for a generic value of the topological mass
$\mu\ne3/(4l)$,\footnote{In what follows we exclude the cases
$\mu=\pm3/(4l)$ since they have triple multiplicity and were already
considered, see Eqs.~(\ref{eq:Fmu=+3/4l_tmC}) and
(\ref{eq:Fmu=-3/4l_tmC}).} is
\begin{equation}\label{eq:F0dmC}
F(u,y)=F_1(u)\left(\frac{y}l\right)^{(4l\mu-3)/[2(l\mu-1)]}+
F_2(u)\ln\left(\frac{y}l\right).
\end{equation}
In the limit $\mu\to\pm\infty$, the exponent in the first term above
takes the value $2$ and the associated mode can be eliminated by
coordinate transformations. Besides, since the associated mass
(\ref{eq:mass_dm2C}) behaves like
\begin{equation}\label{eq:2+NMG}
m^2=\frac1{2l^2}+\mathcal{O}\left(\frac1{\mu}\right),
\end{equation}
one recovers just one of the two modes of the critical solution
(\ref{eq:Fm2=+1/2l2}). Moreover, from the mass expression
(\ref{eq:mass_dm2C}), we notice that the Topologically Massive
Gravity limit, $m^2\to\pm\infty$, is achieved for $\mu\to1/l$. In
this limit, the mode associated to the power law in (\ref{eq:F0dmC})
diverges/vanishes and has no analogue in Topologically Massive
Gravity. In contrast, the logarithmic mode survives and then one
recovers the single critical mode of Topologically Massive Gravity
with $\mu=1/l$ \cite{AyonBeato:2004fq,AyonBeato:2005qq}
\begin{equation}\label{eq:Fmu=+1/lTMG}
\left.F_{\mathrm{TMG}}(u,y)\right|_{\mu=+1/l}=
F_2(u)\ln\left(\frac{y}l\right).
\end{equation}

The last example for which double multiplicity arises is the point
\begin{equation}\label{eq:mass_dm3C}
m^2=\frac{\mu}{2l^2(\mu+1/l)},
\end{equation}
where the root $\alpha=2$ becomes double. The corresponding solution
for a generic value of the topological mass, as long as
$\mu\ne-3/(4l)$, is expressed by
\begin{equation}\label{eq:F2dmC}
F(u,y)=F_1(u)\left(\frac{y}l\right)^2\ln\left(\frac{y}l\right)
+F_2(u)\left(\frac{y}l\right)^{1/[2(l\mu+1)]}.
\end{equation}
In the limit $\mu\to\pm\infty$, the power-law contribution in
the second term of (\ref{eq:F2dmC}) can be eliminated by
coordinate transformations and the associated mass
(\ref{eq:mass_dm3C}) behaves as in Eq.~(\ref{eq:2+NMG}). This
limiting case allows to recover the remaining mode of the
critical solution (\ref{eq:Fm2=+1/2l2}) which was absent in the
previous solution. Taking a look at the Topologically Massive
Gravity limit ($m^2\to\pm\infty$) in (\ref{eq:mass_dm3C}) one
notices that it is in correspondence with the limit
$\mu\to-1/l$. Here again, the mode associated to the power-law
dependence in Eq.~(\ref{eq:F2dmC}) diverges/vanishes and has no
analogue in Topologically Massive Gravity. The logarithmic mode
remains untouched and it becomes exactly the single critical
mode allowed for $\mu=-1/l$
\cite{AyonBeato:2004fq,AyonBeato:2005qq}; namely
\begin{equation}\label{eq:Fmu=-1/lTMG}
\left.F_{\mathrm{TMG}}(u,y)\right|_{\mu=-1/l}=
F_1(u)\left(\frac{y}l\right)^2 \ln\left(\frac{y}l\right).
\end{equation}

These cases exhaust all the AdS-wave solutions (\ref{eq:AdSwave})
one finds for the theory (\ref{eq:NMGC}).

\section{Conclusions}

In this paper, we have studied AdS-wave configurations in
three-dimensional massive gravities.

The first model we considered was the New Massive Gravity recently
proposed by Bergshoeff, Hohm, and Townsend, in
Ref.~\cite{Bergshoeff:2009hq}. In addition to the cosmological
constant, this model has a mass parameter $m$, which is the coupling
constant of the higher-curvature terms that supplement the
Einstein-Hilbert action. For this theory, we considered AdS-wave
configurations, which correspond to exact solutions that can be
thought of as waves propagating on AdS$_3$ spacetime of radius $l$.
These AdS-wave solutions are characterized by a function $F(u,y)$,
which describes the profile of the wave, and depends on the retarded
time $u$ and on the front-wave coordinate $y$.

We have exhaustively explored the space of solutions of this
kind and, in particular, we have shown that special features
occur at the critical values $m^2=\pm1/(2l^2)$. At these
points, solutions with logarithmic fall-off in the Poincar\'e
radial coordinate $y$ arise. This resembles what happens in the
case of Topologically Massive Gravity at the chiral point
$\mu=\pm1/l$ \cite{AyonBeato:2005qq}. At $m^2=+1/(2l^2)$, one
finds that asymptotically AdS$_3$ exact solutions obeying the
weakened fall-off proposed in
\cite{Grumiller:2008es,Henneaux:2009pw} for Topologically
Massive Gravity appear; a fact that is suggested by the
linearized analysis performed in Ref.~\cite{Liu:2009kc} for New
Massive Gravity. This special mass coincides with the point of
the space of parameters at which the central charge of the dual
CFT$_{2}$ vanishes. This naturally leads one to the conjecture
that, likely, the New Massive Gravity of
\cite{Bergshoeff:2009hq} at the point $m^2=1/(2l^2)$ is dual to
a two-dimensional non-unitary conformal field theory if
sufficiently weakened AdS$_3$ asymptotic conditions are
considered. For the range of parameters $m^2>1/(2l^2)$,
however, asymptotically AdS$_3$ solutions obeying stronger
Brown-Henneaux boundary conditions arise, and the theory is
likely unitary.

We also found solutions whose isometry corresponds to the
Schr\"odinger symmetry group. These geometries exist for
$m^2>1/(2l^2)$, and are analogous to those that were recently
considered in the context of the non-relativistic version of
the AdS/CFT duality \cite{Son:2008ye,Balasubramanian:2008dm}.
In particular, the \emph{full} Schr\"odinger symmetry is
achieved for $m^2=17/(2l^2)$.

For all the values $m^2\neq 1/(2l^2)$ the profile function $F$
behaves as a massive scalar excitation, as it satisfies the
Klein-Gordon equation with effective mass
$m_{\mathrm{eff}}^2=m^2-1/(2l^2)$. In fact, the profiles
describe two exact scalar modes sharing the same mass. In
particular, for the case $m^2=-1/(2l^2)$, these modes saturate
the Breitenlohner-Freedman bound for a massive particle on the
AdS$_3$ space where the wave is propagating on.

We also considered the New Massive theory of Gravity coupled to
Topologically Massive Gravity. This introduces a second mass
scale $\mu$ in the theory. For this model, we have exhaustively
explored the space of AdS-wave solutions, and we have shown
that different branches with different asymptotic behaviors
arise. We discussed the effects of turning on the gravitational
Chern-Simons term: This induces a mass splitting that breaks
the mass degeneracy present in the generic modes of the New
Massive Gravity waves. Additionally, several generalization of
the previous results are obtained, for example, \emph{full}
Schr\"odinger invariant backgrounds are obtained now for
$m^2=17\mu/[2l^2(\mu-3/l)]$. We also analyze the different
limits both to New Massive Gravity and to Topologically Massive
Gravity obtaining consistent results. The interplay between the
parity preserving and the parity violating Lagrangians is also
discussed.

A recent paper \cite{ponjas3} studies the linearized solutions of
New Massive Gravity coupled to Topologically Massive Gravity in
AdS$_3$. The asymptotic behaviors of the exact solutions we have
found here realize some of the linearized solutions of
Ref.~\cite{ponjas3}.

\begin{acknowledgments}
The authors are grateful to A.\ Garbarz, A.\ Garc\'{\i}a and J.\
Oliva for useful discussions. This work has been partially supported
by grant 1090368 from FONDECYT, by grant UBACyT X861 from UBA, and
by grants 82443 and 45946-F from CONACyT.
\end{acknowledgments}


\end{document}